%% file: main.tex
\documentclass[conference]{IEEEtran}
 \pdfoutput=1

\ifCLASSOPTIONcompsoc
  \usepackage[nocompress]{cite}
 \else
  \usepackage{cite}
\fi
\ifCLASSINFOpdf
\else
\fi
\usepackage{amsmath,amssymb,amsfonts}

\usepackage{tabularx,booktabs,ragged2e}
\usepackage{commath}
\usepackage{algorithmic}
\usepackage{graphicx}
\usepackage{textcomp}
\usepackage{subcaption}
\usepackage{soul}
\usepackage{xcolor}
\usepackage{enumitem}
\usepackage{multirow}
\usepackage{wrapfig,booktabs}
\usepackage{flushend}

\newcommand{\system}{the system}

\newcolumntype{L}{>{\RaggedRight\arraybackslash}X}
\newcolumntype{C}[1]{>{\centering\arraybackslash}p{#1}}
\usepackage[numbers]{natbib}

\begin{document}
%
\title{Designing Just-in-Time Detection \\for Gamified Fitness Frameworks}

\author{\IEEEauthorblockN{ Slobodan~Milanko, Alexander~Launi, and Shubham~Jain}
\IEEEauthorblockA{{Department of Computer Science} \\
{Old Dominion University}\\
Norfolk, VA, USA \\
{ \it smila001@odu.edu, alaun001@odu.edu, jain@cs.odu.edu}}
}

\IEEEtitleabstractindextext{%
\begin{abstract}
This paper presents our findings from a multi-year effort to detect motion events early using inertial sensors in real-world settings. 
We believe early event detection is the next step in advancing motion tracking, and can enable just-in-time interventions, particularly for mHealth applications. Our system targets strength training workouts in the fitness domain, where users perform well-defined movements for each exercise, while wearing an inertial sensor. We collect data for 20 exercises across 12 users over 26 months. We propose an algorithm to detect repetitions before they end, to allow a user to visualize movement derived metrics in real-time. We further develop a gamified approach to display this information to the user and encourage them to perform consistent movements. Participants in a feasibility study find the gamified feedback useful in improving their form. Our system can detect repetition events as early as 500 ms before it ends, which is 2x faster and more accurate than state-of-the-art trackers. We believe our approach will open exciting avenues for tracking, detection, and gamification for fitness frameworks.
\end{abstract}

 \begin{IEEEkeywords}
 Movement mechanics, Fitness tracking, Wearable sensors, Inertial measurement unit
 \end{IEEEkeywords}
}

\maketitle

\ifCLASSOPTIONcompsoc
    \ifCLASSOPTIONconference
        \IEEEdisplaynontitleabstractindextext
   
    \else
        \IEEEraisesectionheading{\section{Introduction}\label{sec:introduction}}
    \fi
\else
    \IEEEdisplaynontitleabstractindextext
    \section{Introduction}
    \label{sec:introduction}
\fi
\input{introduction.tex}

\input{background.tex}
\input{method.tex}
\input{experiments.tex}

\input{evaluation.tex}
\input{related.tex}

\input{lessons.tex}
\input{discussion.tex}

\ifCLASSOPTIONcaptionsoff
  \newpage
\fi



%
\bibliographystyle{IEEEtran}
\bibliography{sample-bibliography,exercise,Exercise2,mypublications}

%








\end{document}

%% file: introduction.tex

With the proliferation of wearable devices transforming the mobile health space, we are in the midst of a fitness revolution.
This paper presents our findings from a multi-year effort in designing and developing an early event detection (EED) algorithm for movement traces captured from a wearable inertial measurement unit (IMU). By early detection, we intend to detect an event after it starts, as soon as possible, but before it ends. EED has applications in mobile health and across the movement performance spectrum, from personalized athletic training to fine-tuning movement in clinical populations and elite dancers. Such an algorithm can be very useful in supporting just-in-time (JIT) interventions, particularly for injury prevention in the mHealth space. It creates the ability to react to events in real-time, much like humans do, and provide feedback. 

We build and validate our system in real-world environments to detect events in the strength training (or weight training) domain. 
We target strength training because it involves widely accepted movements that are often monitored by a human (a coach or a trainer) who provides corrective feedback in real-time. We detect the movements, typically referred to as repetitions, as a user performs them.
Most event detection algorithms are trained to detect complete events. Therefore, a major challenge in EED is to recognize partial events as they occur in real-time. What makes it more challenging is that the start of the event is unknown and often follows noisy observations. To address these challenges, we develop an online segmentation and EED algorithm that can identify a repetition before it ends.

While motion tracking using inertial sensors has been studied extensively~\cite{closinggap,mouseintheair,armtrak,bringingIOT}, early detection of events is an understudied problem.
Most existing research in the field of movement science relies on dedicated equipment setups, such as 10-15 camera motion 
capture~\cite{mocap,Vicon_2019} systems.
These setups are elaborate and expensive for a regular fitness or physical therapy center, let alone an individual user. 
More recently proposed wearable devices~\cite{Beast2016,athos,PushBand} can track repetitions and sets in a workout. Some of these~\cite{Beast2016} can also compute velocity and strength. However, these approaches have large detection delay and do not provide information on important factors pertaining to user's form. 
Additionally, these solutions rely on labeled training data and often require the user to perform a `proper' movement, which is 
counter-intuitive. Helping the user execute proper movements is not a requirement, but a goal of our system.



Recently, gamification has emerged in the fitness domain to promote health and wellness. It is a trend that uses games or game-like interfaces for influencing behavior, and refers to the use of game design elements in non-game contexts~\cite{Deterding2011}, also called {\it serious games}. Games are often used to incentivize users to achieve their goals. We use game design in this paper, to merge quantifiable  performance metrics with game-like elements to encourage users to perform consistent movements. Our gamified feedback approach is an exploration in understanding user needs and analyze feedback measures.

To design and build this system, we:
\begin{itemize}[leftmargin=*,topsep=0pt]
    \item Conduct a 26 months study to collect data for 20 exercises across 12 users in real-world settings.
    \item Develop an early event detection algorithm for detecting targeted movements in real-time.
    \item Compare our system with state-of-the-art movement trackers to demonstrate that our system provides the lowest delay in event detection.
    \item Design, develop, and evaluate a gamified approach to allow users to visualize  movement derived metrics.
\end{itemize}

%% file: background.tex
\section{Background and Challenges}
\label{sec:background}

Fitness activities can involve a diverse set of movements. For this study, we focus on a specific type of fitness training, often referred to as strength or weight training. Strength training poses a number of unique challenges not present in other forms of exercise like running or cycling.

\subsection{Background}

{\bf Primer to strength training.}
Strength training is a type of physical exercise that focuses on building muscle, strength, and increasing mobility. Throughout this work we use the terms weight training, resistance training, and strength training interchangeably. Workouts are accomplished by completing well-defined movements, called repetitions. Each repetition involves two phases: a lifting (concentric) phase and a lowering (eccentric) phase. The order of the phases may reverse based on the exercise. One or more repetitions performed consecutively without taking a break, are referred to as a set~\cite{hatfield1993hardcore}. Each workout is made up of several sets, with recovery periods or rest times in between.

\noindent {\bf Why strength training as a target application.}
\label{subs:difficulties}
The number of injuries in fitness training is 
staggering~\cite{gray2015epidemiology, hedrick2008weightlifting, andrew2014impact,kerr_2010}. The National Electronic Injury Surveillance System keeps records of the 
common injuries for which people visit the hospital~\cite{cpsc}. In the last five years, there was a total of $547,252$ instances of injury related 
to weightlifting. The trend continued in 2017, where $111,435$ instances were reported. Weight training has a high entrance barrier due to the risk of serious 
injury caused by incorrect motion~\cite{matthews2014beyond, goertzen1989injuries,gray2015causes}. Studies~\cite{faigenbaum2010pediatric,lubetzky2009prevalence} have shown that most injuries are caused by improper lifting techniques, poorly chosen training loads, or lack of qualified supervision.
Strength training and related micro movements are a largely understudied area in the realm of mobile health. Most commercially available products are incapable of tracking these movements in time to provide any kind of feedback or intervention~\cite{BioStrap,Fitbod,FitBit,BarSensei,Beast2016,PushBand,Tendo}. 
We posit that monitoring movement mechanics precisely and providing JIT interventions can have a significant impact on preventing injuries that may be avoidable.

\noindent {\bf Understanding metrics.}
A user's physical form during exercise is most often used to indicate correctness of motion. Typically, form is observed and assessed via visual inspection by a human (an instructor or trainer) leading to subjective and varying assessments of the same movement. Through an extensive literature review~\cite{fleck2014designing, matthews2012cardio, matthews2014beyond, hatfield1993hardcore, bryant2013bench, brzycki1993strength, schoenfeld2015effect}, conversations with fitness instructors, and our collaborators in the Exercise Science department, we compiled a list of carefully selected measurable metrics that can quantify a user's form and performance over time.
With a growing trend in velocity-based training, which posits that covering the same distance quicker can lead to improving power, {\it velocity} has emerged as an important measure of performance. The {\it range of motion} and {\it duration} of each phase (tempo training) affect the velocity, and can be significant even when looked at independently. Another important indicator of form is the {\it stability} of the upper arm or the elbow tracking angle. 
It has been demonstrated that the onset of fatigue can be identified by examining these movements for sticking points.
{\it Sticking points} occur when a user struggles with the resistance. 
Different exercise movements focus on different metrics, and therefore we aim to capture quantifiable movement-related metrics. 
It has been shown that accurately capturing these metrics relies on the ability to measure and track micro movements precisely, and can be performed using an arm-mounted IMU. Details regarding these metrics and their accuracy are discussed in our previous research~\cite{milanko}.

\subsection{Challenges}
\label{sec:challenges} 

{\bf Domain-specific challenges.}
Accurately monitoring athletes' performance based on their motion is still largely an unsolved problem. This is due to a large number of variables involved in determining not only their form and performance, but also overtraining and fatigue leading to potential injuries. The complex nature of various muscle involvement and body movement varies from one exercise motion to another, rendering any modeling extremely challenging.
Strength training athletes perform targeted motion which often have a very small range of motion. Even the slightest of deviations from that movement can lead to injuries. State-of-the-art inertial motion tracking solutions exhibit median errors of 12-15 cm~\cite{closinggap}. This error is very large compared to the linear movement in some of the exercises. To address this challenge, our key insight is that IMUs can track rotations more accurately than linear movement, and most body movements can be represented as a series of rotations. 


\noindent {\bf Real-time implementation challenges.}
One of the most significant challenges in motion tracking is the ability to do it in real-time. This is extremely important for real-world applications, particularly those targeted at injury prevention, where timeliness is crucial. For enabling effective JIT interventions and to measure motion-related metrics, the system needs to accurately identify: (1) when each set begins and ends (2) when each repetition begins and ends. Sets of exercise repetitions are often interspersed between body movements performed during recovery which generates significant noise in the data. Most existing solutions~\cite{PushBand,Beast2016,BarSensei} require the user to manually mark the beginning and end of each set. Although effective, we believe that this increases the cognitive burden on the athlete and can prove to be a distraction or annoyance over time. Most of these solutions do not perform early event detection, and can only detect a repetition when the next repetition is in progress. This delay renders them impractical for enabling any real-time feedback. We address this challenge by designing an early event detection system that can detect a rep before it ends, allowing the user to receive feedback on their motion as they perform it.

\noindent {\bf Challenges in feedback mechanisms.} 
While training, the cognitive abilities of most athletes are compromised, and any feedback must be broken down to simplify its interpretation and assimilation. Since many athletes are not exercise-science domain experts, we cannot expect prior knowledge of technical jargon. Moreover, the feedback given by instructors is often short sentences that are corrective or affirmative statements about the user's form. Since an acceptable response is to imitate human behavior, it makes the mode of feedback an open research question. Our solution to this challenge is to gamify the feedback provided to the user. Gamification in the fitness domain has proven to be effective for other forms of exercise, such as running~\cite{zombiesrun}. 

%% file: method.tex
\section{Early Event Detection}
\label{sec:eed}

\begin{figure}[!t]
    \centering
    \includegraphics[height=0.3\textwidth]{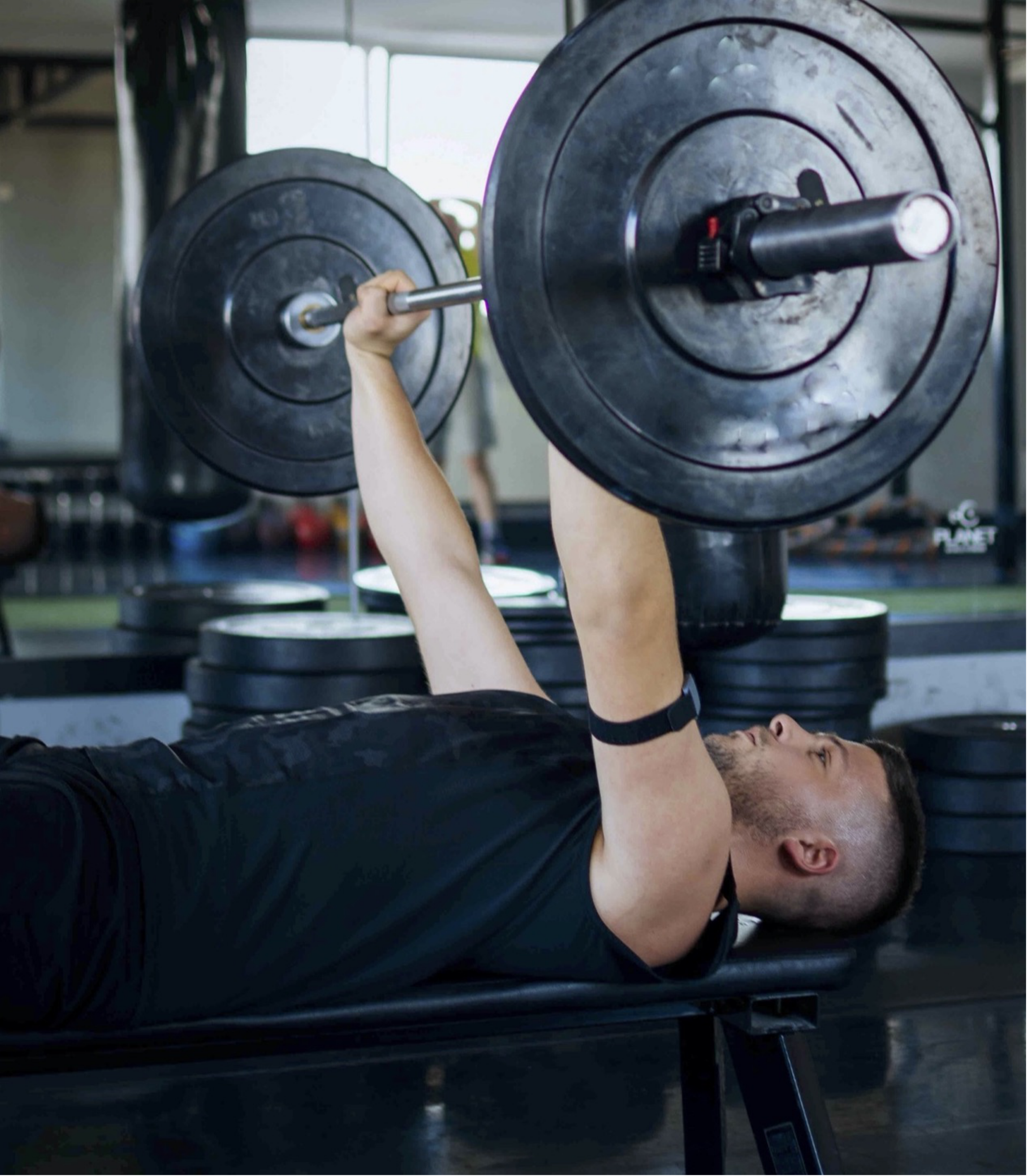}
    \caption{A user performing a bench press with the sensor attached to their upper left arm.}
    \label{fig:sensor}
\end{figure}

We study how to enable early detection of a movement in real-time, with partial data. Our review of the literature yielded very limited solutions for early event detection~\cite{dorschky2015,wang2018,hoai2012}, where most of them were focused on computer vision techniques. While offline event detection in temporal sequences has been studied extensively, early detection is largely understudied. Early detection is different from forecasting, because forecasting predicts the future whereas early detection derives information from partial traces obtained in real-time. Time-series forecasting generates raw data for the future observations, but does not detect the event class for the present and past data. Conceivably, forecasting could be a first step in such detection, but it is likely to add more sources of error in event detection. Our system processes a temporal sequence of orientations from an inertial measurement unit (IMU) in real-time. The IMU is mounted on a user's right arm, as shown in Figure~\ref{fig:sensor}. Repetitions performed during training are the events of interest. In this work we use the terms {\it event} and {\it repetition} interchangeably. We believe that by detecting events early we can derive measurements and actionable insights from motion. 

\subsection{Design Considerations}

To generate actionable information via early event detection in real-world conditions we define the following design goals:

\textbf{Scalable.} {\it Minimize the number of sensors.} 10-15 camera motion capture (MoCap) systems~\cite{Vicon_2019} and RGB-Depth cameras are often used by researchers in biomechanics for tracking human movement mechanics precisely. By using a single wearable sensor to monitor movement mechanics, we develop an approach that is scalable and can be used widely at low-cost.

\textbf{Translatable.} {\it Track variety of movements.} We ensure that the proposed algorithms work across a wide variety of movements. We gathered data for 20 different upper body movements to validate the performance of our system. Unlike many machine and deep learning approaches we minimize the dependency on a large volume of data. By tracking movement mechanics and employing probabilistic models for EED we can bootstrap our approach for new exercises with minimal data.

\textbf{Unobtrusive.} {\it Minimize user interaction.} In contrast to existing solutions, we aim to minimize user's interaction with the system interface. Our system segments users' motion into constituent sets and repetitions without requiring any input from the user.

\textbf{Real-time.} {\it Provide actionable insights with minimum delay.} Most users are used to receiving feedback from instructors in real-time as they workout, allowing them to correct their movements during the exercise. To this end, our system processes the stream of IMU data in real-time to identify transitions from noise to signal (at the beginning of a set) and detects events (repetitions) as they happen.

\subsection{Online Set Segmentation}

\begin{figure}[!t]
    \centering
    \includegraphics[width=0.485\textwidth]{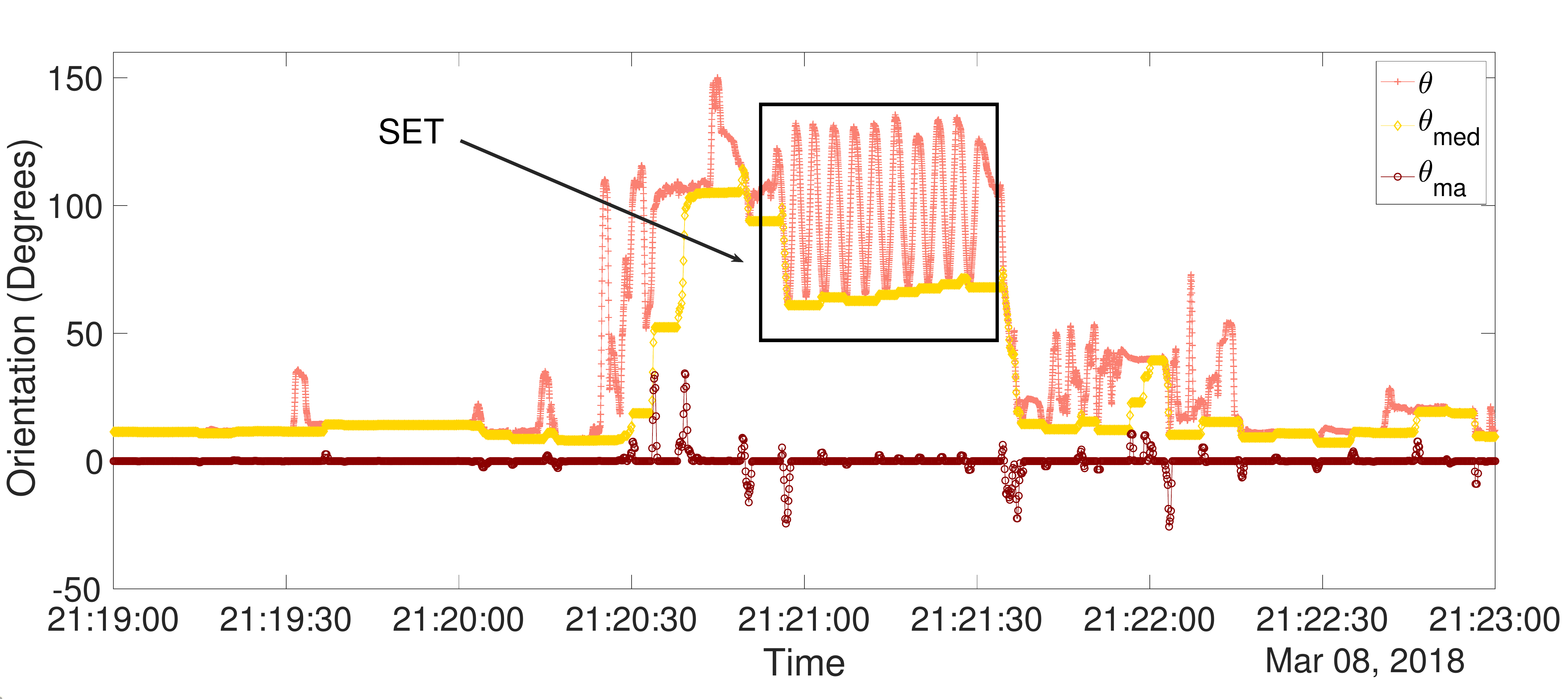}
    \caption{Processing arm orientation ($\theta$) through a cascade of filters to detect the onset of sets.}
    \label{fig:angles}
\end{figure}

\textbf{Sensor Placement and Data Preprocessing.} For this study we mounted the IMU on the user's upper arm, as shown in Figure~\ref{fig:sensor}. We utilized the sensor placement and shoulder coordinate system developed in our previous work~\cite{milanko}. The orientation of the arm at any given time $t$, $\theta_t$ is calculated by fusing data from the accelerometer and gyroscope using a Kalman Filter.  $\theta$ is the time-series signal that denotes the orientation of the user's arm over time, used to identify the various phases of a user's movement. A sample orientation trace is shown in red in Figure~\ref{fig:angles}.

\noindent \textbf{Detecting set onsets.} A primary challenge in supporting unobtrusive approaches is to determine when events of interest happen. Accurately identifying these events requires the ability to segment noise arising from motion in between sets that may not be relevant to the motion under study. For our system, we're interested in determining when a user is doing repetitions and when they are not. Multiple repetitions are typically conducted together as a set. Figure~\ref{fig:angles} shows that a set is significantly different from other data, but the onset of a set can be very noisy as a user prepares to perform the exercise. As a first task we focus on determining the beginning and ends of each set in real-time. We refer to each set as the movement of interest (MOI).

We process $\theta$ through a cascade of filters. For the online segmentation, the device running the algorithm (a smartphone) captures five samples, $s$, over Bluetooth. In the first filtering operation, we smooth $s$ by passing them through a median filter, with a 100 sample window size, to retain the macro movements. This low-pass filtering operation on the incoming samples is performed in context of previously received samples. 
A higher order median filter preserves large changes in the signal, thus removing minute movements, jitters, and sporadic movements. 
The result of this operation, $\theta_{med}$, is shown in Figure~\ref{fig:angles}.
We cascade the median filter with a moving average filter operation. This filter further smooths the resultant signal, $\theta_{ma}$ and improves the signal-to-noise ratio. 
Figure~\ref{fig:angles} illustrates that $\theta_{ma}$ captures the sharp transitions within the signal. Transitions into MOI are marked by a positive correlation between $\theta_{med}$ and $\theta_{ma}$, and a negative correlation for transitions out of the MOI.

We compute the range, $R$, in a 5 step window of $\theta_{ma}$. 
and utilize a data mining technique, a covering algorithm, to determine a dynamic threshold for $R$. This technique can be viewed as a rule processing algorithm that aims to cover a vast majority of positive samples in a given data set~\cite{witten2016data}, here being when MOI begins and ends. The threshold is given by Equation~\ref{eqn:dynamicthreshold}: 
\begin{equation}
    I = \frac{1}{2n}\left (\sum_{i=1}^n{x_i}\right )
    \label{eqn:dynamicthreshold}
\end{equation}
where $x$ denotes $\theta_{med}$. When $R$ exceeds this dynamic threshold, we identify it as the beginning of a set and the repetition detection module is triggered.  

\subsection{Early Repetition Detection}

Once \system~detects the onset of a set, the next step is to identify each repetition event. Each repetition exhibits a V-like structure and can be further split into two phases, lifting and lowering, as discussed in Section~\ref{sec:background}. Our objective is early identification of each repetition.

\begin{figure}[!t]
    \centering
    \includegraphics[width=0.42\textwidth]{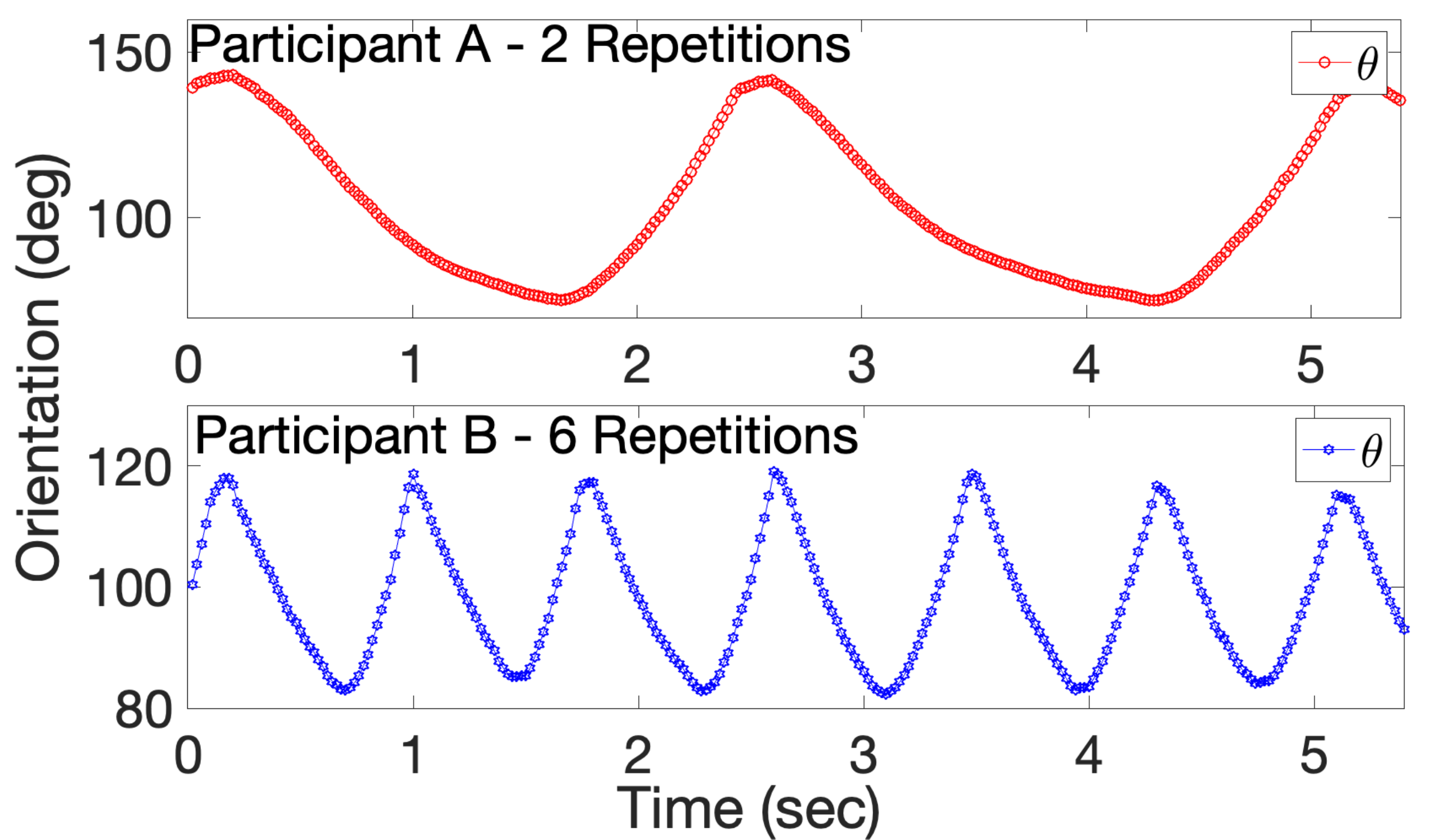}
    \caption{Variance in tempo and arm orientation for 2 participants performing the bench press. The top subplot demonstrates a participant completing 2 repetitions, while the bottom shows another participant completing 6 repetitions in the same amount of time. Note the difference in their range of motion.}
    \label{fig:tempoVar}
\end{figure} 

We use a sliding window on the signal $\theta$. 
Due to the variability of strength training repetitions, both in terms of tempo (duration) and arm orientation (changes in $\theta$), we aggregate the sliding window dynamically, instead of keeping it a fixed size~\cite{bao2004activity}. 
An example of such variability in tempo can be observed in Figure~\ref{fig:tempoVar}. Different participants take different amount of time to perform their reps, thus a fixed window size for data processing can significantly limit the perspective.
We start with a window of length $w$.
The window $\theta_{t-w:t}$ is matched using dynamic time warping (DTW) against a template, $temp$, which is common for all exercises. If two neighboring windows conform with $temp$, with a distance lower than an empirically determined threshold, the windows are aggregated. Note that, lower distance between signals denotes a closer match. This process continues until incoming windows no longer match $temp$, at which point $w$ is reset to its initial value. 

For each window, we compute various attributes (observations), such as correlation, standard deviation, range, and gradient, and use them to determine whether or not the user is performing a repetition. To model this we use a Dynamic Naive Bayes (DNB) model~\cite{learningDNB,bove2019}, which has the functional form: 

\begin{equation}
    P(A,S) = P(S_1)\prod_{t=1}^{T - 1} P(S_{t + 1}|S_t) \prod_{t = 1}^{T} \prod_{m=1}^{M} P(A_{t}^m | S_t)
\end{equation}
where, $P(S_1)$ is the prior probability of being in state $S_1$ at time $t=1$, $P(S_{t + 1}|S_t)$ is the transition probability between $S_t$ and $S_{t + 1}$, and $P(A_{t}^m | S_t)$ is the probability of the observed feature $m$ at time $t$ given the state $S_t$. DNB models are an extension of Hidden Markov Models (HMM) and are particularly advantageous to model a system with multiple observation symbols. As shown in Section~\ref{sec:eval}, multiple observations are critical to the robustness of our system.
Our system is said to be in one of two possible states - {\it Event} or {\it Non-Event}, based on whether or not that window belongs to a repetition. 
The prior probability, transition matrix, and observation matrix are estimated from a separate dataset of about $400$ repetitions, and the parameters are learned using the Baum-Welch algorithm.
In every step, the system makes four observations, given by $A_t$, where 
$A_t$ = \{$corr_t, rom_t, std_t, gradient_t$\}.
$corr$ measures the correlation between $\theta_{t-w:t}$ and $temp$. $rom$ represents change in orientation within the current window, $std$ the standard deviation, and $gradient$ the negative and positive slope distribution in the window $\theta_{t-w:t}$. We discretize each of these continuous observations into a common encoded space - low, medium, and high. 
As the algorithm processes incoming samples of data the sliding window starts aggregating incrementally at the beginning of a repetition. As more data becomes available the DNB model indicates a transition between states which determines if a user is currently performing a repetition.

\begin{figure}[t]
\centering
   \begin{subfigure}{0.23\textwidth}
    \includegraphics[scale=0.088]{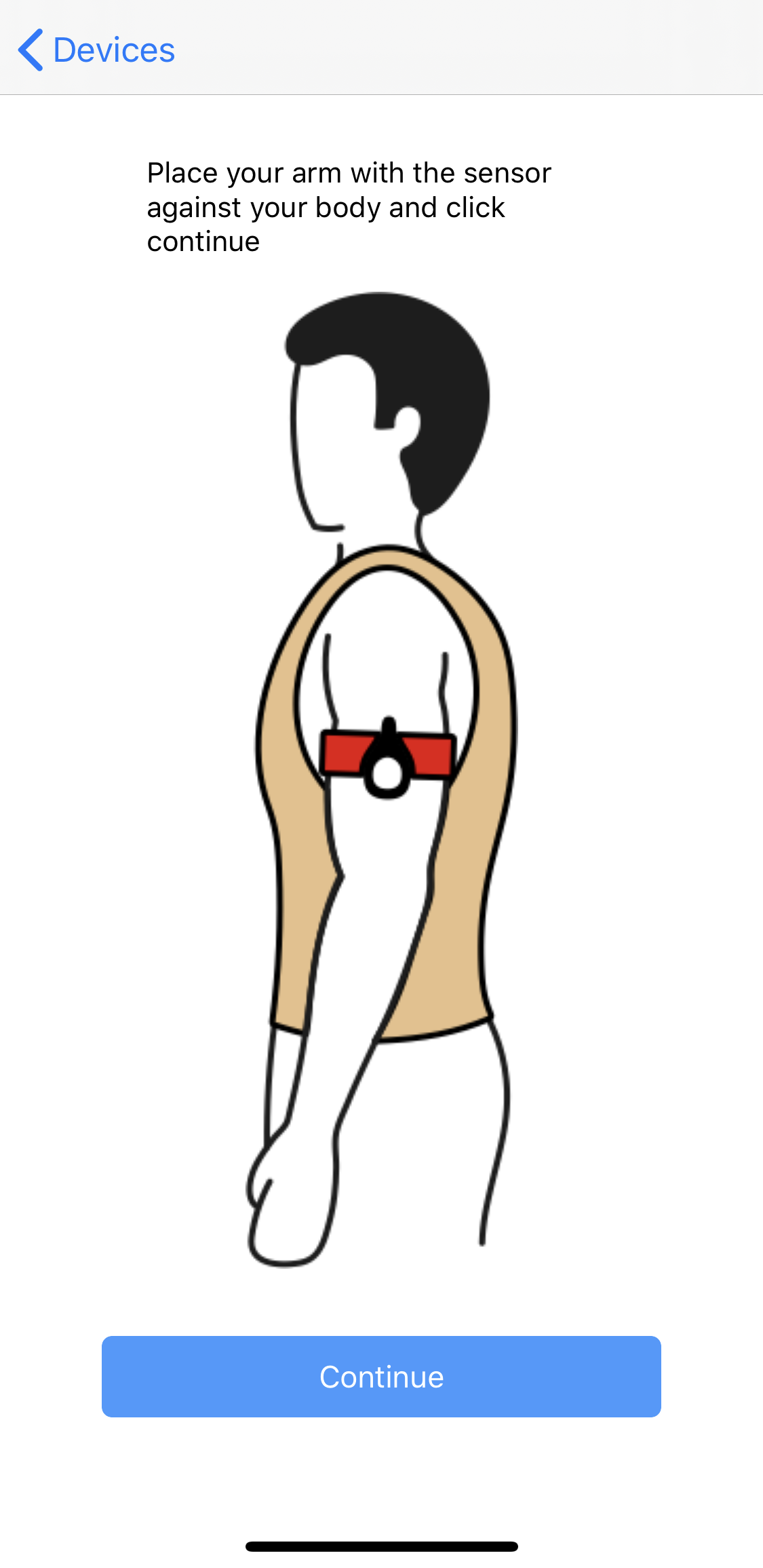}
    \caption{Home Screen}
    \label{fig:liftright_home}
   \end{subfigure}
   \begin{subfigure}{0.23\textwidth}
    \includegraphics[scale=0.088]{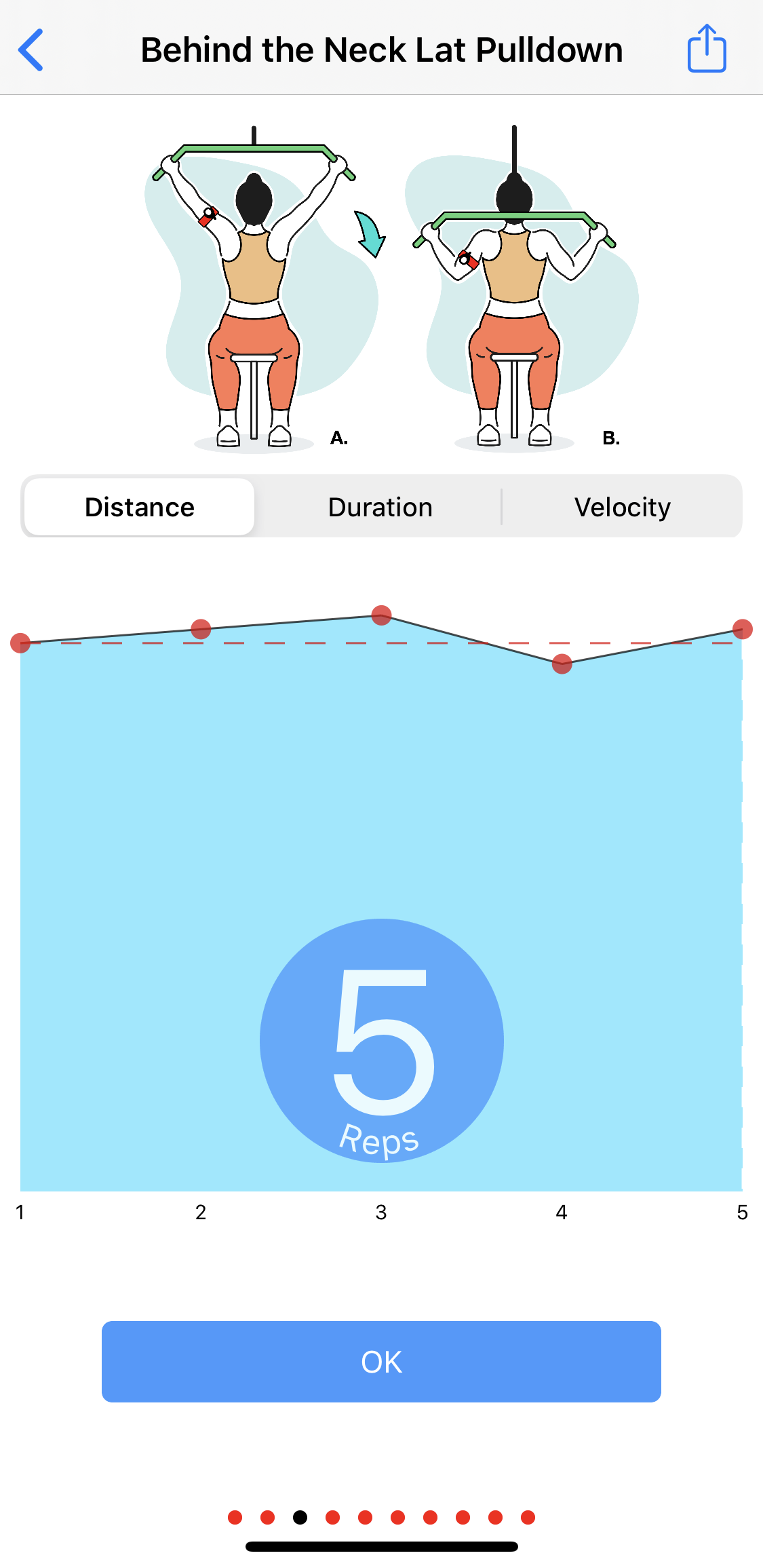}
    \caption{Connect-the-dots Screen}
    \label{fig:liftright_exercise}
   \end{subfigure}
    \caption{Screenshots from our prototype application.}
\end{figure}

\subsection{Gamified Feedback for JIT Detections}

Gamification has the potential to improve quality of learning through enhanced engagement with the user. In the fitness domain, gamification has demonstrated promising outcomes~\cite{zhao2016gamified, goh2015gamification, hassan2019motivational}. Until now, game mechanics have been used for macro-level motion, such as active minutes, distance walked, ran, or bicycled. Often, they also rely on social networking to entice friends and family to compete with each other for rewards (often points)~\cite{challengesapp, garminchallenges}. For most existing applications, the focus is on improving high-level performance. We have explored the use of game-based visual dynamics for tracking movements mechanics and allowing the users to view it in real-time.  

\textbf{Game Requirement.}
The primary goal of this game is to provide users with a visualization of their form. To this end, timeliness of this display is very important, making low latency a requirement of the system. 
The game is designed for users at all levels. To avoid cognitive overload, we focus on displaying three metrics: range of motion, velocity, and duration, and a user can choose to view any one of these three metrics while exercising.
A major challenge is that in the absence of guidelines, most users cannot determine correctness of form by simply viewing their data in real-time. Keeping this in mind, the game provides users with a baseline to encourage consistency in movements. In our current prototype this baseline is derived from the user's first repetition movement. It is important to note that this rep is not used to train the model, just to provide a visual baseline. We believe that with a visual baseline, users will be encouraged to meet the baseline with every movement, even as fatigue sets in. If a user thinks they might not be able to meet the baseline for the next movement, it might also encourage them to take a break. This is significant, since our findings (detailed in Section~\ref{sec:lessons}) show that most users continue to perform repetitions even when their form is deteriorating.

\begin{table}[!t]
\small
 \begin{tabularx}{0.49\textwidth}{@{} l *{3}{L} c @{}} 
 \hline
 Exercise & \#~Events & \#~Days  & Time (min) & \#~Subjects\\
 \hline\hline
  Bench Press & 1558 & 30 & 541  & 12 \\
 \hline
  Lateral Pulldown & 1485 & 23 & 330 & 11  \\
 \hline
 Behind Neck Press & 1353 & 23 & 398 & 10  \\ 
 \hline
  Battle Rope & 1109 & 6 & 68 & 5 \\
 \hline
   Tbar Row & 323 & 4 & 29 & 4 \\
 \hline
   Sit Ups & 247 & 3 & 28 & 2 \\
 \hline 
   Push Ups & 231 & 4 & 43 & 2 \\
 \hline
 Flys & 207 & 6 & 55 & 4   \\
 \hline
  Stiff Arm Pulldown & 155 & 4 & 45 & 3 \\
 \hline
  Barbell Row & 146 & 4 & 37 & 3 \\
 \hline
  Lateral Rise & 143 & 6 & 35 & 5 \\
 \hline
 Kettle-bell Swing & 131 & 2 & 29 & 1 \\
 \hline
 Pull Ups & 93 & 2 & 36 & 1  \\
 \hline
  Upright Row & 89 & 2 & 21 & 1 \\
 \hline
 Push Jerk & 76 & 2 & 29 & 2  \\
 \hline
  Dips & 72 & 3 & 40 & 3 \\
   \hline
 Shoulder Press & 59 & 2 & 13 & 1  \\
 \hline
  Front Rise & 50 & 1 & 12 & 1 \\
 \hline
  Squats & 48 & 1 & 25 & 1 \\
 \hline
  Rear Delt Row & 48 & 1 & 11 & 1 \\
 \hline
\end{tabularx}
\caption{Total recorded events (repetitions) and sessions (one per day for non-consecutive days) per exercise.}
\label{table:collectedData}
\end{table}

\begin{figure}[t]
\centering
    \includegraphics[width=0.45\textwidth]{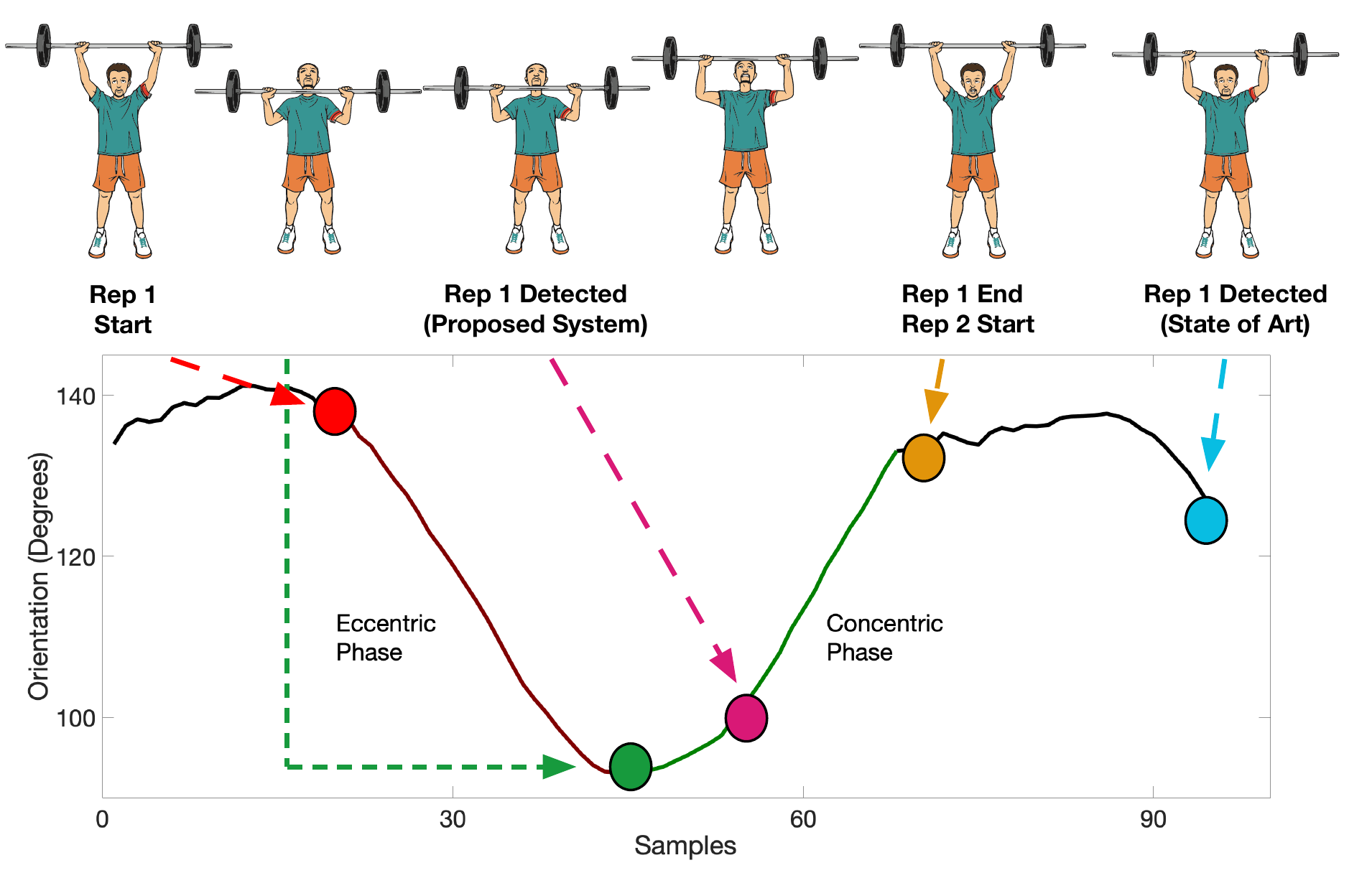}
    \caption{Early event detection: Illustration of when a repetition for overhead press can be identified via the proposed system and the state of the art. }
    \label{fig:digitalRepRepresentation}
\end{figure}

\begin{figure*}[t]
    \centering
    \includegraphics[width=\textwidth]{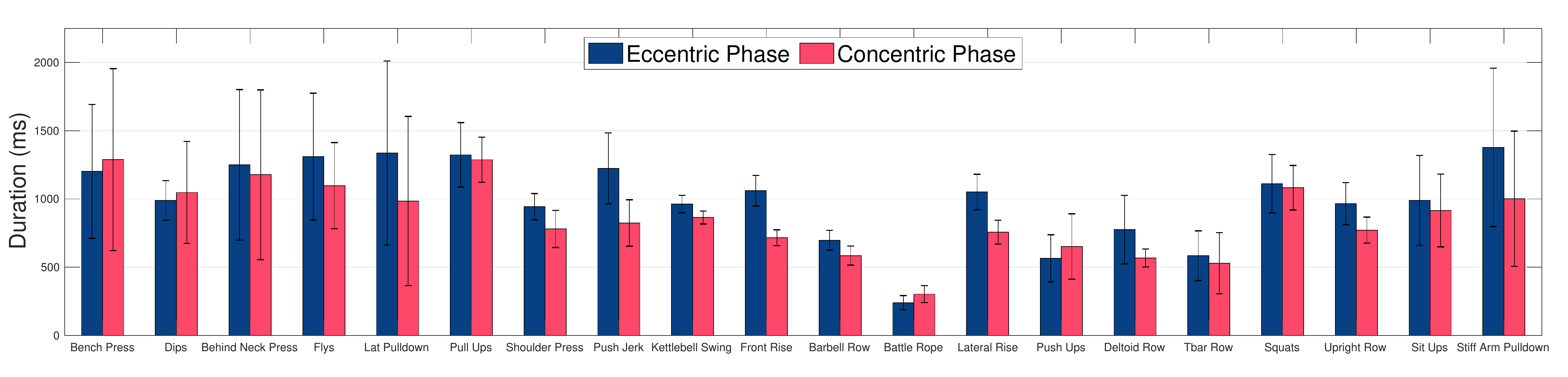}
    \caption{Average eccentric and concentric phase duration for 20 strength training exercises. The errors bars represent standard deviation. }
    \label{fig:concentricEcceentricPhaseDurations}
\end{figure*}

\begin{figure*}[!t]
\begin{subfigure}{0.478\textwidth}
    \includegraphics[width=\textwidth]{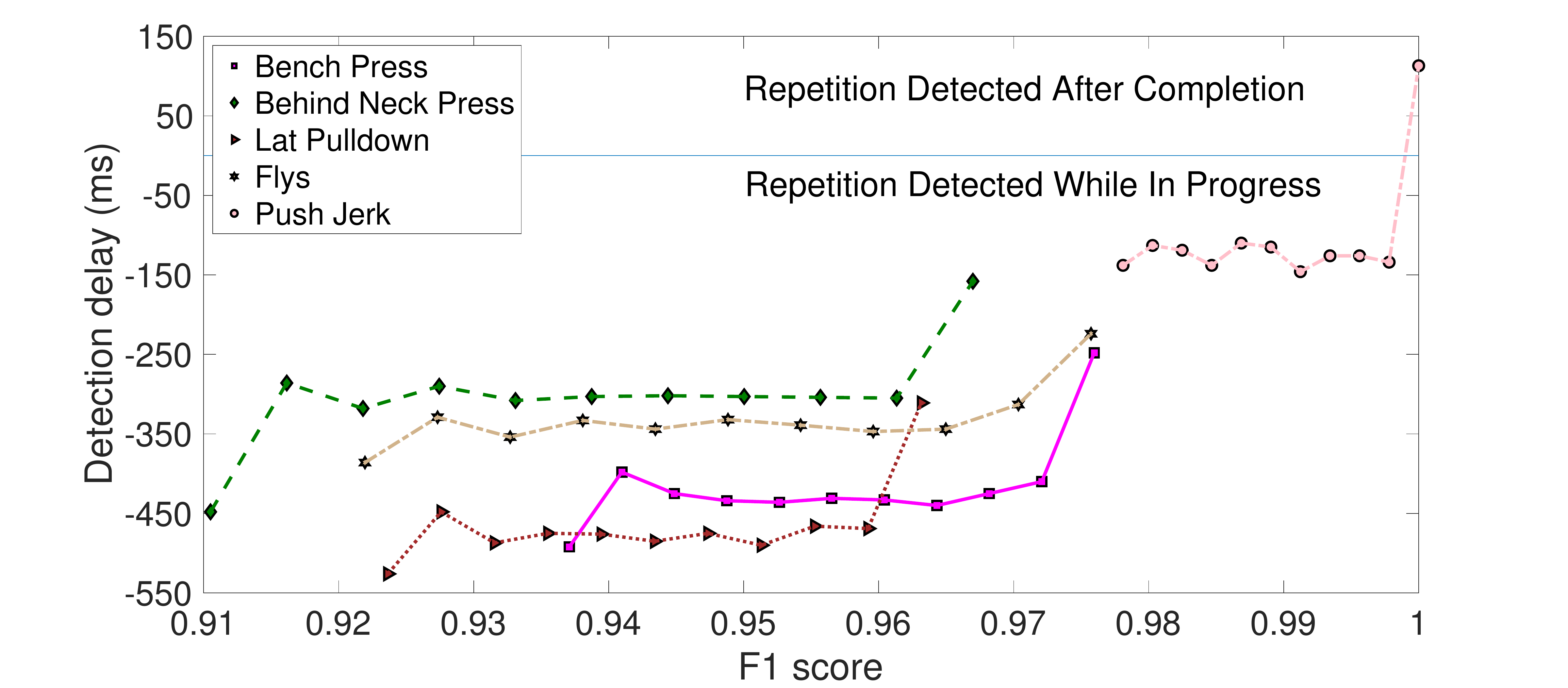}
    \caption{Trade-off between early repetition detection accuracy and delay.}
    \label{fig:delayModelEvolution}
\end{subfigure}
\begin{subfigure}{0.475\textwidth}
    \includegraphics[width = \textwidth]{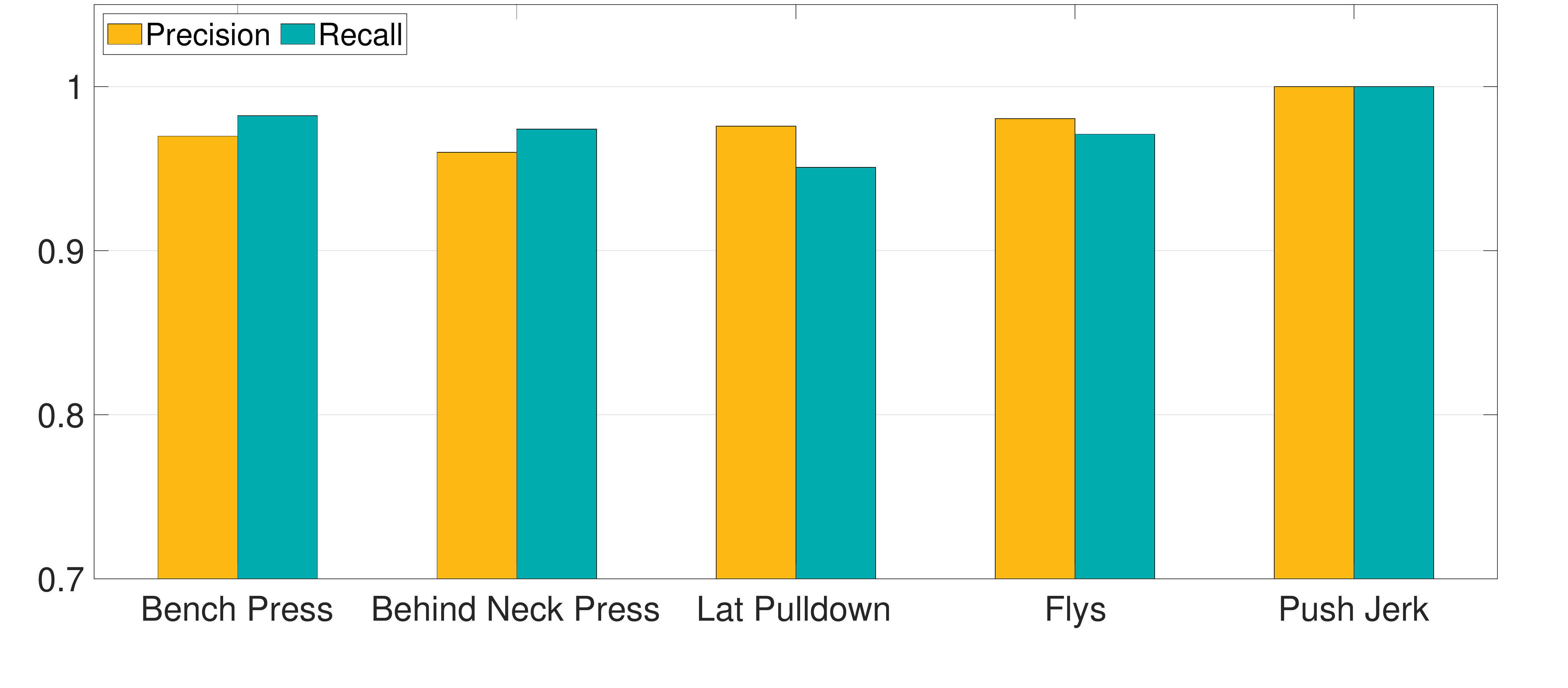}
    \caption{Precision-Recall for repetition identification.}\label{fig:precisionRecall}
\end{subfigure}
\caption{System performance for 5 strength training exercises: Bench Press, Behind Neck Press, Lat Pulldown, Flys, and Push Jerk.} \label{fig:pr}
\end{figure*}

\textbf{Game Design and Implementation.}
The algorithm for early event detection, detailed in Section~\ref{sec:eed}, is implemented in an iOS app. The home screen, shown in Figure~\ref{fig:liftright_home}, shows the user how to mount the sensor. Upon pressing continue, 10 exercises are available to the user to select from. The user can swipe between screens to change exercises. For each exercise, the display allows the user to pick from one of three metrics: Distance (range of motion), Duration (tempo), and velocity. We do not want to overwhelm the user by displaying all the metrics at the same time. Since users are accustomed to training styles that may focus on one metric more than the others, the ability to choose between them gives them the flexibility to continue training the way they typically do. The screen for a single exercise is shown in Figure~\ref{fig:liftright_exercise}, with the distance metric selected. As the user performs repetitions, the rep number is updated in real-time. For each rep, a dot appears on the screen displaying the metric for that rep. The position of the dot on the screen is mapped to the user's arm position in real-time. The first rep is used to create a baseline (dashed red line in Figure~\ref{fig:liftright_exercise}). The tracking is implemented as a {\it connect-the-dots} game. The data is also logged for offline viewing by the user.
The game design is based upon observations from our previous work~\cite{milanko}, which shows that user's form and performance (range of motion, velocity, tempo, elbow stability) deteriorate significantly within each set. 
By encouraging the user to connect the dot for each rep along the baseline, we aim to assist the user in maintaining consistency among repetitions.
The algorithm automatically stops tracking when the set is over, and resumes tracking for the following set. Sets are separated by a vertical line in the connect-the-dots game screen. 
We evaluate our game design for both accuracy and engagement from the user's perspective.
We obtained feedback from our users after every session, and discuss our findings in Section~\ref{sec:lessons}. 

%% file: experiments.tex
\section{Experiments and Data Collection}


To determine set segmentation and repetition identification efficiency we tested against the collected strength training data. Table~\ref{table:collectedData} details the exercises captured, along with the number of repetition events, sessions, and participants for each. Sessions for the exercises were conducted once per day. Keeping the safety of the subjects in mind no same exercise was performed on consecutive days, and appropriate recovery time recommendations were followed. 
Overall our data contains $7,623$ repetitions. In an IRB-approved study, the experiments were conducted with 12 volunteers, out of which 9 were male and 3 female, ranging between the ages of 27-45. At the beginning of each session, an IMU was placed snug around the upper arm of the participant. A laptop was used to connect to the sensor via Bluetooth Low Energy (BLE), capturing accelerometer and gyroscope data at $50Hz$. Once participants were ready a video recording was captured with Logitech and Ximea MC023CG-SY-UB cameras at 120+ frames per second. The videos were used as ground truth to annotate the start and end of each phase (lifting or lowering) in each repetition. 
During the session participants worked out at their regular gym, and were encouraged to behave naturally and perform exercises as they usually do. Once an exercise was complete the sensor was removed and the camera capture ended. We simulate the sequential arrival of samples in MATLAB
for design and evaluation of the real-time algorithm.

A widely shared online survey allowed us to understand user behavior and requirements during weight training which informed the design of our real-time application (Section~\ref{sec:lessons}).
The real-time gamified version was implemented on an Apple iPhone using Swift and C as the programming platform. The accelerometer and gyroscope sensors were sampled at $50Hz$ from the MetaWear Motion R platform, through BLE. An iPhone with the app running was mounted in front of the user. 
In a feasibility study with 16 participants, the mobile version was used to determine user engagement, influence, and usability. 4 out of the 16 participants were female and 12 were male. Users were given a brief introduction on how to use the mobile application, the metrics displayed, and the game. They were allowed to experiment with the application before using it. At the end of the session, users were asked to fill a feedback form. The feedback from users is discussed in Section~\ref{sec:lessons}.


%% file: evaluation.tex
\begin{figure}[!t]
\centering
    \includegraphics[width=0.48\textwidth]{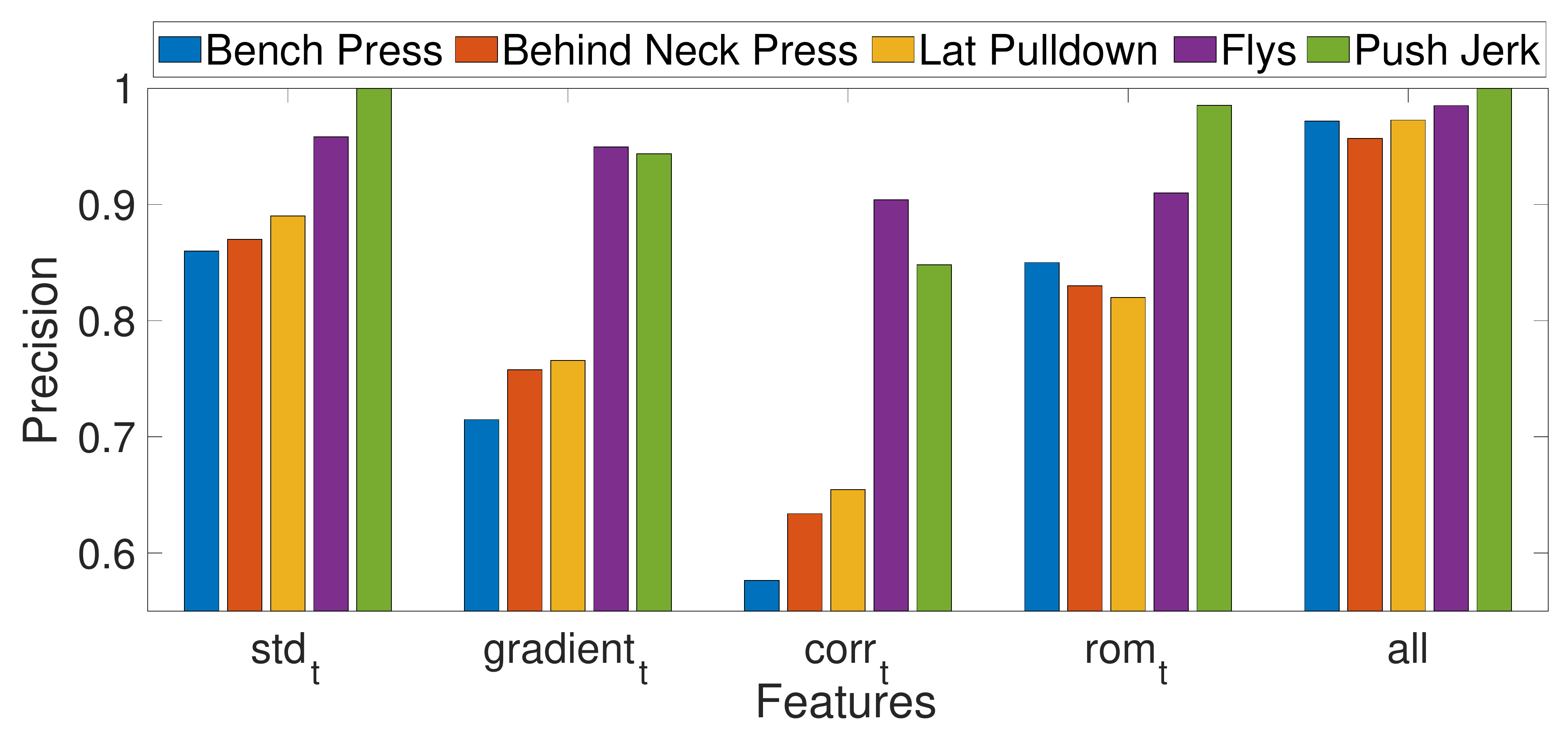}
    \caption{Impact on system precision when using individual features, $std_t$, $gradient_t$, $corr_t$, and $rom_t$, versus all features together, $all$.}
    \label{fig:featureAddition}
\end{figure}

\section{Performance Evaluation}
\label{sec:eval}

\subsection{Event Detection}

\begin{table}[h]
\centering
 \begin{tabular}{l|c} 
 \hline
    Exercise &  Accuracy\\ \hline
    Bench Press & 99.93\% \\
    Behind Neck Press & 98.94\% \\
    Lat Pulldown & 99.79\% \\
    Flys & 92.98\% \\
    Push Jerk & 89.13\% \\
\end{tabular}
\caption{Segmentation accuracy.}
\label{tab:setsegacc}
\end{table}

\textbf{Set Segmentation.} Videos captured during data collection were used as ground truth. We manually annotated the start and end of every set, repetition, and phase within the repetition. Table~\ref{tab:setsegacc} shows the accuracy for identifying sets in a real-time temporal sequence. Note that set segmentation denotes identifying transitions from noise to signal, and vice versa. Noise is comprised of the movements performed by users when they are not exercising. These include stretching, drinking water, and using a towel, which are often performed during recovery time in between sets. We can see that our system achieves very high accuracy for identifying the beginning and end of sets. Bench press exhibits highest accuracy due to the unique position the body is in, i.e. lying down. Push jerk suffers in accuracy due to the explosive nature of the movement.
Out of total 626 minutes of noise, 502 minutes were successfully removed.

\noindent \textbf{Repetition Identification.}
Figure~\ref{fig:concentricEcceentricPhaseDurations} shows the average phase duration of the 20 strength training exercises we collected data for. It shows the variability among various exercises, and across participants for each exercise. We can see that often a repetition phase lasts for less than $1-1.5$ seconds, making detection and timeliness requirements very rigid.
Figure~\ref{fig:digitalRepRepresentation} shows the point in a repetition when our system detects it as compared to state of the art trackers.
Commercial trackers detect a repetition after it is over and the following repetition has started, which is too late to generate actionable feedback. Our system is capable of detecting repetitions as they enter the second phase of the repetition. This early detection, combined with computation of movement-derived metrics, is capable of providing real-time feedback regarding the ongoing repetition.

In evaluating our system's ability to identify repetitions we investigate the trade-off between delay and accuracy for our EED algorithm. 
Figure~\ref{fig:delayModelEvolution} displays this trade-off, where accuracy is represented by the F1 score. For clarity only data for five exercises is shown. A detection delay of $0$ (marked by the horizontal line in Figure~\ref{fig:delayModelEvolution}) indicates that a repetition was detected when it ended. A positive delay means that the repetition was detected after it ended, and a negative delay represents early detection of the rep. 
It can be seen that as more data from a repetition is available, detection delay increases, and the system can identify events with higher accuracy.
Figure~\ref{fig:delayModelEvolution} also demonstrates that early detection is possible for a variety of exercise movements, even for those that are performed in bursts like push jerk. Our system can identify repetition events as early as $550$ ms before they end, with an accuracy higher than $90\%$ for all 5 exercises. Figure~\ref{fig:precisionRecall} shows the precision-recall values for the same set of 5 exercises. Our repetition detection algorithm exhibits high precision and recall for a variety of exercises.

Figure~\ref{fig:featureAddition} shows system precision when using individual features in Dynamic Naive Bayes model. We can see that $std_t$ and $rom_t$ are effective in detecting repetitions early, but their performance varies across exercises. Combining all 4 features yields a consistent performance of higher than $95\%$ precision across workouts and lends itself well to scalability to more exercises.



\begin{figure}[!t]
    \includegraphics[width=\columnwidth]{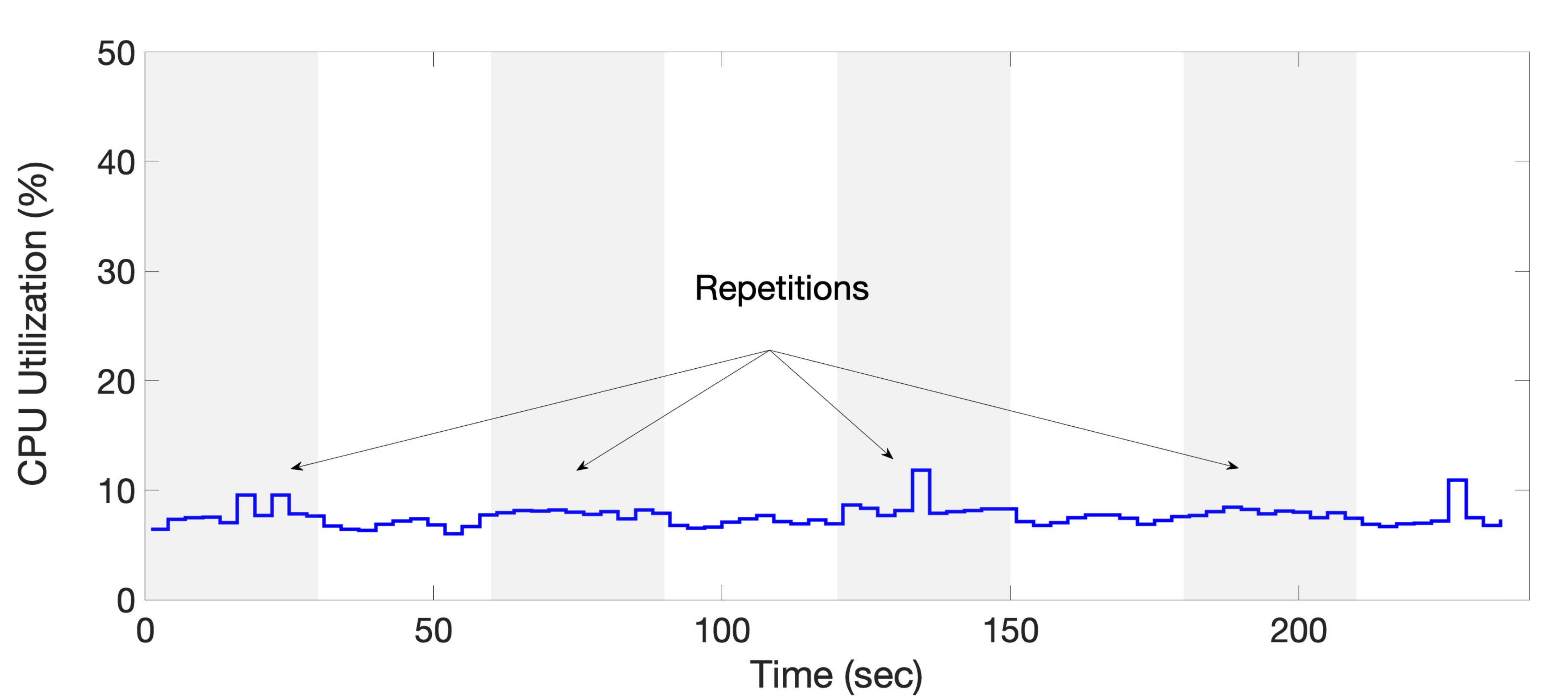}
    \caption{CPU utilization for our prototype app. Shaded areas represent time user was actively performing repetitions.}
    \label{fig:cpuvis}
\end{figure}

\noindent \textbf{Resource Consumption.}
Our mobile app's performance was profiled using Apple's profiling and performance monitoring tool, Instruments. Apple provides an event logging API \texttt{Signpost} which was used to map changes in our model's state to resource utilization. The shaded areas in Figure~\ref{fig:cpuvis} indicate when the user was performing repetitions of an exercise. On an iPhone X running iOS 13.3.1 Instruments indicated that our app's {\it energy impact} was from {\it very low} to {\it low}. CPU utilization for the lifetime of the application ranged from a minimum of 6.0\%, to a maximum of 14.8\%, and average CPU utilization was 7.5\%. Data shown is averaged across 2 runs performing 30 seconds of repetitions followed by 30 seconds of rest performed 4 times. CPU spikes did not correspond with any specific process, and were not reproducible across runs. Maximum CPU spike observed was 14.8\%, which did not have an impact on overall system performance. It is evident that our segmentation algorithm optimizes resource consumption, keeping it low when the user is not exercising.


\subsection{Comparison with state-of-the-art}

\begin{figure}[!t]
    \includegraphics[width = \columnwidth]{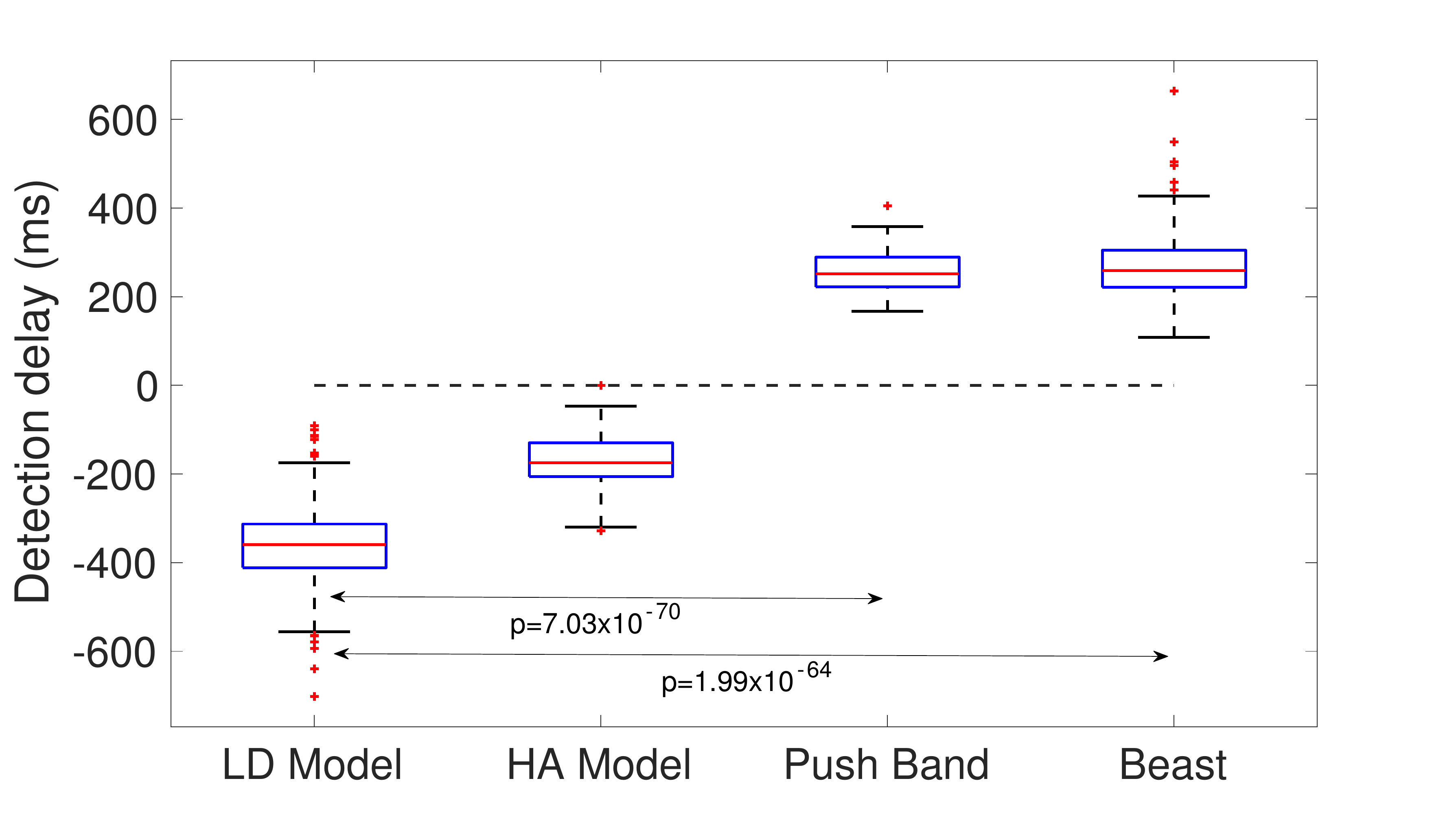}
    \caption{Detection delay for a Low Delay (LD) Model, High Accuracy (HA) Model, and commercial tools - Push Band and Beast. Negative delay indicates that a repetition was identified before completion. }\label{fig:delayComparisonWithCommercial}
\end{figure}

We created two models to test against two commercial products, Push Band and Beast, in four Shoulder Press sessions. The Low Delay (LD) Model is tuned for early detection and the High Accuracy (HA) Model for higher accuracy. 
A collection of 400 repetitions from three participants was used to perform the comparison, with the eccentric phase of the repetitions averaging to 800 ms in duration and concentric phase 550 ms.

During exercise, the HA model exhibited a 1\% false positive rate, Push band 2\%, LD model 4\%, and Beast 29\%. False negatives were only observed in Beast and our LD model, at 0.3\% and 0.07\% respectively. While none of the devices/algorithms were able to identify a repetition during the first phase, both proposed models performed early detection during the second phase of a repetition. 
A comparison of the detection delay for both commercial trackers and two models proposed by us can be observed in Figure~\ref{fig:delayComparisonWithCommercial}. The horizontal line with a detection delay of 0 indicates the time when a repetition ends. Beast and the Push Band identified repetitions after they were complete. 87\% of the repetitions identified by the Push Band and 72\% of those identified by Beast were detected after the following repetition started. 
Our Low Delay model detected repetitions with only 33\% of the data from the second phase, as early as 650 ms before repetitions ended. The High Accuracy model detected repetitions when 60\% of the second phase was complete, up to 300 ms before they ended. 





%% file: related.tex
\section{Related Work}
\label{sec:related}

\textbf{Inertial Motion Tracking.} 
IMUs have long been used for motion tracking, due to their affordability and availability within mobile systems. While commonly used for localization~\cite{ndjeng2011low,LookUp,ped_hotPlanet,vehicularsensing} and event detection~\cite{Chang:2007:TFE:1771592.1771594,predriveid}, their use to identify events from partial traces is largely understudied.
Inertial measurement units also suffer from poor accuracy caused by system bias and drift. Shen et al.~\cite{armtrak} demonstrate how measuring displacement in free form activities causes massive errors. This is also true in strength training~\cite{beckham2019reliability, flores2016validity}. As a solution to this problem, sensor fusion and stochastic modeling techniques have been used to design fusion models for improving accuracy~\cite{yean2016algorithm, armtrak,closinggap, lou2011sensor, wang2015mole}.

\noindent \textbf{Exercise Tracking.}
Perhaps the most accurate tools in exercise tracking today are computer vision~\cite{appenrodt2010data, ali2011gait,scattolin2012comparison, khurana2018gymcam} and pulley systems~\cite{Tendo}. While they are predominant in research and professional settings, high operational cost and infrastructure support deter public adoption. Alternatively, low cost and consumer familiarity of wearable devices are making single and multi sensor wearables more popular~\cite{dong2007real,de2008real,pan2013intelligent}. They have been used for tracking joints during physical rehabilitation~\cite{pan2013intelligent, raso2010m, lin2013online,liftright_demo}, performance and fatigue in strength training~\cite{milanko, pernek2013exercise, morris2014recofit}, and even sports analysis~\cite{mannini2013activity,Chang:2007:TFE:1771592.1771594,justin_wearsys}. To our knowledge however, most real-time wearable solutions trade delay and high user engagement for accuracy. This includes obtrusive and supervised event segmentation~\cite{PushBand, Beast2016, BioStrap, raso2010m, bittel2016accuracy, pernek2013exercise, spina2013copdtrainer}, multi-second delay in receiving feedback~\cite{pruthi2015maxxyt, bittel2016accuracy}, and requiring users to remain idle after events are complete for assessment~\cite{BarSensei}.

\noindent \textbf{Feedback in exercise monitoring.}
Bove~\cite{bove2019} investigated the effect that the "quantified self" had on patient engagement in a medical setting. They found that although patients were wearing smart devices, many did not remain engaged with the data over time. Work by Jarrahi et al.~\cite{jarrahiFitbit} showed that unless data continues to be valuable to users they will stop using the device.   Zhao et al.~\cite{zhao2016gamified} developed an iOS game which used exercise duration as input. They demonstrated that goal-based games using wearable sensors as input devices can be motivating factors for encouraging people to exercise.
In a follow-on long-term study it was shown that a gamified approach can have a positive impact on users' exercise habits~\cite{zhao2016game2}.



%% file: lessons.tex
\section{LESSONS LEARNED}
\label{sec:lessons}

\subsection{Determining user needs and feedback}

\begin{table}[!t]
\small
 \begin{tabularx}{0.49\textwidth}{l|c}
 \hline
    \multicolumn{2}{C{0.47\textwidth}}{Q: I would like the following measured during my workout}\\ \hline
    Number of sets & 62\% \\    
    Number of reps in each set & 68\% \\
    Total time spent lifting & 41\% \\
    Can I do another rep & 35\% \\ \hline
    \multicolumn{2}{C{0.47\textwidth}}{Q: I would wear a fitness tracker if it calculated all the metrics I selected}\\ \hline
    Definitely yes or Probably yes & 68.5\%\\
    Undecided & 15\%\\
    Definitely not or Probably not & 16\% \\
\end{tabularx}
\caption{Summary of responses from online survey.}
\label{tab:surveyresponse}
\end{table}
\textbf{Assessing user needs via online survey.} To understand the needs of the athletic community, including both - professionals and enthusiasts, we designed an anonymous online survey and shared it widely. The responses of 187 people from 12 countries were analyzed. The survey was designed to understand user behavior with respect to workout routine, to identify the metrics they are interested in, and to gauge acceptability for a technology that provides them access to these metrics.
Table~\ref{tab:surveyresponse} provides a summary of our findings from this survey. More than 60\% users were interested in a system that measures the number of sets and repetitions performed during a workout. They are also interested in recording the duration of their repetitions. Moreover, 68.5\% of the participants expressed willingness to wear a fitness tracker that captures their metrics of interest.
We use the responses from this online survey to guide us in the design goals stated in this paper.


\noindent \textbf{Post-session assessment of gamified feedback.} Post-session user experience questionnaires were conducted for 16 participants in a feasibility study, after each session to gather users' impressions of the gamified interface. Of these, 4 were professional trainers/instructors. A summary of the responses is shown in Table~\ref{tab:poststudy}.
Our respondents overwhelmingly found that the device did not physically impact their workout. 93\% of participants found the device very comfortable and 87\% responded that the system seemed very accurate in capturing metrics. Participants also found the real-time visualization helpful.
A previous study on wearable mHealth devices found that many users were highly concerned with the aesthetics of the device~\cite{10.1145/3300061.3345432}. Participants in our study described our prototype sensor as, "sleek", "modern", and "user friendly".

\noindent \textbf{Discussion.} There are three important takeaways from these studies. First, most people acknowledge the desire to obtain fine-grained exercise-related metrics and do not mind wearing a tracker to that effect. This substantiates our basic premise and positions the proposed system to address this unmet need.
Second, a gamified interface that allows the user to view their metrics with a faster response time as they worked out, elicited positive reactions from the users. Since most participants did not find our gamified interface distracting, this validates the simplicity of our game design which makes it easy for participants to comprehend their metrics while working out.
Third, while people benefit from visualization, two-thirds of our participants prefer some form of audio feedback. A challenge in incorporating audio feedback lies in the limited options for personalization. With the real-estate of a smartphone screen, we can employ visualizations to convey multiple things in a simple manner. However, since most people listen to music while working out, audio feedback can easily blend into their music. Based on this feedback, we're exploring audio-based gamification techniques that look into modifying music tone and tempo to keep users engaged and informed about their form and performance.

\begin{table}
\small
 \begin{tabular}{p{3.7cm}|c|c|c} 
    \hline
     & Very & Somewhat & Not at all \\ \hline
    Did seeing the metrics in real-time help you improve your form? & 74\% & 20 \% & 6\% \\
    Was the visualization distracting to you? & 0\% & 13\% & 87\%\\
    Would you prefer audio feedback? & 20\% & 47\% & 33\%\\
    Was the sensor/arm band comfortable? & 93\% & 7\% & 0\%\\
\end{tabular}
\caption{Responses from post-session questionnaire.}
\label{tab:poststudy}
\end{table}

\subsection{Experiment-driven Insights}

A major goal of our experiments in terms of data collection was to use real-world settings. To achieve this, we allowed users to exercise at their own gyms and keep their regular workout routines. This required us to (1) adhere to each gym's privacy policy and regulations on data/video recording, (2) keep up with all our participants' schedules, and (3) validate our system across equipment from multiple gyms. 
Another interesting observation was that noise caused by random movements in between sets can sometimes look very similar to sets. A specific example is a person who stretched in between sets by still keeping their hands on the bar. Another example is a person who, while doing a bench press, did not get up to recover in between sets. Such movements make it difficult to separate noisy data using simple signal processing techniques.


Our data shows large differences in individual lifting styles. In addition to the duration (shown in Figure~\ref{fig:concentricEcceentricPhaseDurations}), the velocity, range of motion, and other critical metrics that indicate form can vary widely. 
Early on, when we were only collecting data with no visual feedback, users often overemphasized a specific goal in their training, for example performing a pre-decided number of repetitions. With the real-time gamified approach, the same users were paying more attention to their form and taking breaks when needed. This helped them decouple from their habit of performing repetitions past the point of exhaustion, at the cost of deteriorated form. This observation points to opportunities to support JIT interventions to prevent injuries by warning users against performing strenuous movements under fatigue. 


The feasibility study with gamified feedback also allowed us to understand cognitive load during exercise. Most fitness trackers provide simple numerical data during training, but after a while it gets hard for users to process the numbers. Our participants were excited by the intuitive game-like interface that allowed them to track their movements easily.
Some of our participants performed vigorous movements in a deliberate attempt to "break the system".  Others, adjusted the dynamics of their movement to test how well the real-time feedback reflected their motion. 
%
Our proposed techniques proved robust enough to provide consistent performance for diverse users.




\subsection{High-Impact mHealth Applications}

Although many populations could benefit from this technology, we discuss an example clinical application here -
rehabilitative settings for common shoulder pathologies. A major focal point for rehabilitation of patients with shoulder impingement ~\cite{Fees_1998}, anterior shoulder instability~\cite{Fees_1998}, and glenohumeral~\cite{Tucci_2011,Reinold_2009} is proper form and slow, incremental progression from isolated to complex (e.g. bench press and pulldown) exercises. 
The ability to capture movement mechanics to derive form-based performance assessments would allow for the removal of continued guidance by on-site professionals by allowing at-home guidance that could reduce travel time, associated cost, and possibly improve exercise adherence. Moreover, combined with techniques that can monitor physiological parameters for users in motion, such as breathing~\cite{aspiro}, our system can enhance user performance.
Our system would be beneficial for both user/patient and their practitioner as it can provide useful feedback on the patient's progress and form, allowing for enhancement of home-based interventions.

%% file: discussion.tex
\section{Conclusion}

This paper reports our experience in designing, developing, and validating an early event detection approach for data collected over 26 months for 20 different strength training exercises. We use a single wearable inertial sensor that is mounted on a user's arm. We demonstrate that we can detect repetitions as early as $500$ ms before they end, with accuracy higher than $90\%$ for different exercises. 94\% participants in our feasibility study found the real-time gamified feedback useful in improving their form.
We believe that the approach and results presented here can have a significant impact in supporting JIT intervention techniques for mHealth applications. We demonstrate our approach in the context of strength training, but the techniques extend across the movement performance spectrum to dancers, athletes, and clinical populations.